\input{aipcheck}

\documentclass[
    ,final            
  ]
  {aipproc}

\layoutstyle{6x9}

\begin{document}

\title{Late-time acceleration in a brane \\ with curvature effects}


\author{Mariam Bouhmadi-L\'opez\footnote{email:mariam.bouhmadi@fisica.ist.utl.pt}\,\,}{
  address={Centro Multidisciplinar de Astrof\'{\i}sica, Departamento de F\'{\i}sica, Instituto Superior T\'ecnico,\\ Av. Rovisco Pais 1,1049-001 Lisboa, Portugal}
}

\author{Paulo Vargas Moniz}{
  address={Departamento de F\'{\i}sica,
Universidade da Beira Interior, Rua Marqu\^{e}s d'Avila e Bolama,
6201-001 Covilh\~{a}, Portugal}
}


\begin{abstract}
  In this paper we investigate if
the phantom-like regime in the LDGP model can be enlarged  by the inclusion of a
Gauss-Bonnet (GB) term into the bulk. 
However, we show that the opposite occurs: the GB effect seems instead to induce
a breakdown of the phantom-like behaviour at an even smaller redshift.
\end{abstract}

\maketitle


\section{Introduction}

One of the most astonishing discoveries of the last years has been the discovery that the universe has started recently to accelerate. A possibility to describe such an acceleration is within modified theories of gravity like  the $\Lambda$DGP which, in addition, is able to mimic a phantom-like behaviour without invoking any real phantom matter on the brane or in the bulk.

More precisely, the  $\Lambda$DGP scenario is a 5D brane-world model with infrared
modifications to general relativity caused by an induced gravity
term on the brane
\cite{Sahni:2002dx,Lue:2004za,Lazkoz:2006gp,BouhmadiLopez:2007ts}.
The model is based on the normal branch of
the Dvali-Gabadadze-Porrati proposal (DGP)
\cite{Dvali:2000hr}, being the brane
filled with cold dark matter (CDM) and a cosmological constant which drives
 the late-time acceleration of the brane. The phantom-like behaviour
 is a consequence of the extra-dimension which screens the
  brane cosmological constant  and it is based in mapping the brane evolution  to that of an equivalent
4D general relativistic phantom energy model
\cite{Sahni:2002dx,Lue:2004za}.
  More precisely, the basis of this mimicry is an effective
  energy density (in the 4D general relativistic picture) corresponding to the cosmological constant
  corrected by the curvature effect due to the induced gravity
  term on the brane. This effective energy density grows as  the
brane expands and therefore effectively it behaves as a phantom
fluid; i.e. $w_{\rm{eff}}<-1$, where $w_{\rm{eff}}$ corresponds
to the ratio between the effective energy density and the effective
 pressure\footnote{
 Other brane proposals aiming to produce such a mimicry
 are based on a bulk filled with matter and/or on an energy
 exchange between the brane and the bulk, therefore modifying
 the effective equation of state of dark energy on the
 brane. \cite{BouhmadiLopez:2005gk}}

The effective description of the phantom  behaviour in the $\Lambda$DGP model  breaks down at
a finite redshift; i.e. the effective energy density
 vanishes and becomes negative over a certain
redshift. When the effective energy density vanishes, the effective equation of state blows up. Given that
the phantom-like behaviour results from ($i$) induced gravity effects on
the brane causing curvature corrections and ($ii$) describing the brane model as a 4D relativistic phantom energy setup, could  the break down of the phantom-like behaviour be eliminated by considering
further curvature effects on the brane-world scenario? This is the
main question we address in this paper. We will model such additional and new curvature effects through a
Gauss-Bonnet term (GB) in the bulk \cite{Kofinas:2003rz,Bouhmadi2008}.

\section{The model}

We consider the $\Lambda$DGP brane, i.e. the normal DGP branch filled with CDM and a cosmological constant, with GB corrections in the bulk. Then the Friedmann equation reads \cite{Bouhmadi2008}
\begin{equation}
E^2(z)=\Omega_m(1+z)^3+\Omega_\Lambda - 2\sqrt{\Omega_{r_c}}\left [1+\Omega_\alpha E^2(z)\right ]E(z),
\label{Friedmannz}
\end{equation}
where we have assumed a mirror symmetry across the brane, $E(z)=H/H_0$ and
\begin{equation}
\Omega_m=\frac{\kappa_4^2 \rho_{m_0}}{3H_0^2},\,\,\,\, \Omega_\Lambda=\frac{\kappa_4^2 \Lambda}{3H_0^2},\,\,\,\, \Omega_{r_c}=\frac{1}{4r_c^2H_0^2},
\end{equation}
are the usual convenient dimensionless parameters while the new parameter $\Omega_{\alpha}$ is defined as
\begin{equation}
\Omega_{\alpha}=\frac{8}{3}\alpha H_0^2.
\end{equation}
Finally, $r_c$ is the crossover scale in the DGP model and $\alpha$ is the GB parameter in the bulk action. 

Evaluating the Friedmann equation (\ref{Friedmannz}) at $z=0$
gives a constraint on the cosmological parameters of the model
\begin{equation}
\Omega_m+\Omega_\Lambda
=1+2\sqrt{\Omega_{r_c}}\left(1+\Omega_{\alpha}\right).
\label{cosmoconstraint}
\end{equation}
This constraint implies that the region $\Omega_m+\Omega_\Lambda<1$ is unphysical.
Moreover, although the brane is spatially  flat, the previous constraint can be interpreted
in the sense that our model constitutes a mimic of a closed FLRW universe. Furthermore, by imposing that the universe is currently accelerating, we obtain another constraint
on the set of cosmological parameters $\Omega_m$, $\Omega_{r_c}$ and $\Omega_\alpha$, which reads
\begin{equation}\label{constraintacce}
3\Omega_m<2+2(1+\Omega_\alpha)\sqrt{\Omega_{r_c}}.
\end{equation}
On the other hand, it can be shown that the brane never super-accelerates; i.e. the Hubble rate decreases as the brane accelerates.

\section{Phantom-like behaviour on the brane}

The phantom-like behaviour on the brane
 is based in defining a corresponding  effective energy density
$\rho_{\rm{eff}}$ and an effective equation of state with parameter
$w_{\rm{eff}}$. More precisely, the effective description
is inspired in writing down the modified Friedmann
equation of the brane as the  usual relativistic
Friedmann equation so that $H^2=
\frac{\kappa_4^2}{3}(\rho_m+\rho_{\rm{eff}})$ where
  $\rho_{\rm{eff}}$ reads
\begin{eqnarray}
\rho_{\rm{eff}}
&=\frac{3 H_0^2}{\kappa_4^2}\left[\Omega_\Lambda-2\sqrt{\Omega_{r_c}}(1+\Omega_\alpha E^2(z))E(z)\right].
\label{rhoeff}
\end{eqnarray}
This effective energy density corresponds to a balance between the cosmological constant and geometrical effects encoded on the Hubble parameter. On the other hand, gravity leakage at late-time screens the cosmological constant like in the $\Lambda$DGP scenario \cite{Sahni:2002dx,Lue:2004za}. This phantom-like behaviour is obtained without any matter violating the null energy condition and without any super-acceleration on the brane.

As in  the $\Lambda$DGP model,
it is possible to define an effective
equation of state or
parameter $w_{\rm{eff}}$ associated to  the effective energy density as
\begin{equation}
\dot\rho_{\rm{eff}}+3H(1+w_{\rm{eff}})\rho_{\rm{eff}}=0.
\end{equation}
This effective equation of state is defined in analogy with the standard relativistic case.
Then using Eq.~(\ref{rhoeff}), we obtain
\begin{eqnarray}
1+w_{\rm{eff}}
=\frac23\frac{\dot H/H_0^2\sqrt{\Omega_{r_c}}\left(1+3\Omega_\alpha E^2(z)\right)}{E(z)\left[\Omega_\Lambda-2\sqrt{\Omega_{r_c}}\left(1+\Omega_\alpha E^2(z)\right) E(z)\right]}.\nonumber
\end{eqnarray}
Because the brane never super-accelerates,
i.e.  $\dot H <0$,
we can then conclude that $\rho_{\rm{eff}}$
mimics the behaviour of a phantom energy component on the brane:
I.e. $1+w_{\rm{eff}}<0$,
as long as the effective energy density $\rho_{\rm{eff}}$ is positive.
However, there is always a redshift $z_b$ at which $\rho_{\rm{eff}}$ vanishes
 and therefore the phantom mimicry ceases to be valid.  The above behaviour raises the following possibility:
Can the phantom behaviour break down in our model at a redshift $z_b$
, such that $z_b>\tilde{z}_b$, where $\tilde{z}_b$ is the redshift at which the phantom mimicry is no longer valid in the $\Lambda$DGP model? 
The answer will depend on the cosmological parameters
that characterise both models, three
 in the $\Lambda$DGP scenario namely
($\Omega_{\tilde m}$,$\Omega_{\tilde\Lambda}$,$\Omega_{\tilde{r}_c}$)
and four in the model we are analysing namely
($\Omega_m$,$\Omega_\Lambda$,$\Omega_{r_c}$,$\Omega_\alpha$).

\begin{figure}
  \includegraphics[width=1.0\textwidth]{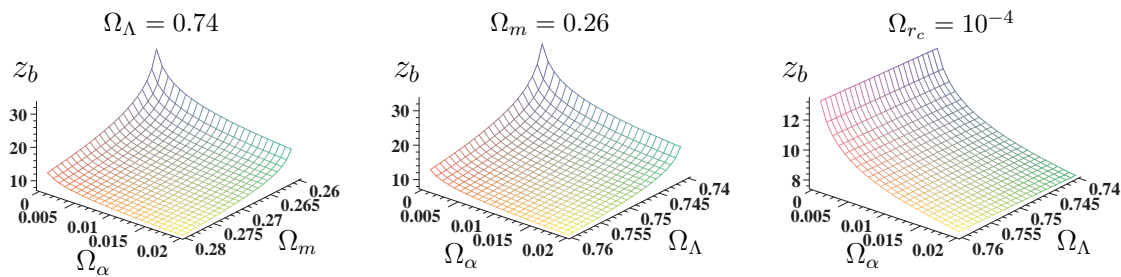}
  \caption{Plot of the redshift  $z_b$ at which the phantom-like behaviour in the $\Lambda$DGP-GB model breaks down. When $\Omega_{\alpha}=0$, we obtain the redshift  $\tilde z_b$ at which the phantom-like behaviour in the $\Lambda$DGP scenario breaks down. As can be noticed
$z_b<\tilde z_b$.}
\label{zbbreak}
\end{figure}

We consider that only one of the parameters ($\Omega_\Lambda$, $\Omega_m$, $\Omega_{r_c}$) is fixed in the $\Lambda$DGP-GB brane. We further assume that the given value of the fixed parameter is very close to  that obtained by constraining the $\Lambda$DGP model with
observational data ($H(z)$, CMB shift parameter and SNIa data)\cite{Lazkoz:2007zk}
\begin{equation}\label{eq60}
\Omega_m=0.26\pm0.05, \quad \Omega_{r_c}\leq 0.05.
\end{equation}
Our results are presented in figure \ref{zbbreak}:  The phantom-like behaviour in the $\Lambda$DGP-GB brane breaks
down sooner (smaller redshift) than in the $\Lambda$DGP brane. 

\section{conclusions}

We have analysed the $\Lambda$DGP-GB model which corresponds to a 5D brane-world model where the bulk is
a 5D Minkowski space-time. The model contains  a GB term in the bulk and an induced gravity term on the brane. Our analysis was performed for the normal branch which we have assumed to be
filled by CDM and a cosmological constant, the latter driving the
late-time acceleration of the brane. We have shown how the brane
accelerates at late-times. 

On the other hand, we have shown that there is a phantom-like behaviour on the brane without resorting  to any real phantom matter on the brane or the bulk. However, the phantom-like behaviour in the $\Lambda$DGP-GB brane breaks
down sooner (smaller redshift) than in the $\Lambda$DGP brane.  Namely, the $\Lambda$DGP has a regular
phantom-like behaviour for $[0, \tilde{z}_b)$ whereas  the phantom-like behaviour
in the $\Lambda$DGP-GB model
 is regular only for $[0,z_b)$, with $z_b<\tilde{z}_b$.
Thus,  we conclude that the GB term does extend the regime of validity of the phantom mimicry in the
$\Lambda$DGP model.

\begin{theacknowledgments}
MBL is  supported by the Portuguese Agency Funda\c{c}\~{a}o para a Ci\^{e}ncia e
Tecnologia through the fellowship SFRH/BPD/26542/2006.
This research work was supported by the grant
FEDER-POCI/P/FIS/57547/2004.
\end{theacknowledgments}

\end{document}